\documentclass{article}
\usepackage{epsf,psfig,gcdv}
\def\reference{\parskip 0pt\par\noindent\hangindent 0.5 truecm}

\def\spose#1{\hbox to 0pt{#1\hss}}
\def\simlt{\mathrel{\spose{\lower 3pt\hbox{$\mathchar"218$}}
     \raise 2.0pt\hbox{$\mathchar"13C$}}}
\def\simgt{\mathrel{\spose{\lower 3pt\hbox{$\mathchar"218$}}
     \raise 2.0pt\hbox{$\mathchar"13E$}}}

\def\apj{ApJ}
\def\apjs{ApJS}

\def\aap{A\&A}

\def\prl{Phys.Rev.Lett.}

\def\mnras{MNRAS}
\def\aj{AJ}
\def\physrep{Phys. Rep.}

\begin{document}

\small
\shorttitle{$^6$Li from Virialization Shocks}
\shortauthor{T. K. Suzuki \& S. Inoue}
%
%
\title{\large \bf
Cosmic Ray Production of $^6$Li by Virialization Shocks 
in the Early Milky Way}

\author{\small 
 Takeru K. Suzuki$^{1}$ and Susumu Inoue$^{2}$
} 

\date{}
\twocolumn[
\maketitle
\vspace{-20pt}
\small
{\center
$^1$Dept. of Physics, Kyoto University, 
Kyoto, 606-8502, Japan; 
stakeru@tap.scphys.kyoto-u.ac.jp, \\
$^2$Max-Planck-Institut f\"ur Astrophysik, 
Karl-Schwarzschild-Str. 1, Postfach 1317, 85741 Garching, Germany
\\[3mm]
}

%
\begin{center}
{\bfseries Abstract}
\end{center}
\begin{quotation}
\begin{small}
\vspace{-5pt}
The energy dissipated by virialization shocks
during hierarchical structure formation of the Galaxy 
can exceed that injected by concomitant supernova (SN) explosions.
Cosmic rays (CRs) accelerated by such shocks
may therefore dominate over SNe in the production of $^6$Li
through $\alpha + \alpha$ fusion without co-producing Be and B.
This process can give a more natural account of the observed $^6$Li abundance
in metal-poor stars compared to standard SN CR scenarios.
Future searches for correlations between the 
$^6$Li abundance and the kinematic properties of halo stars
may constitute an important probe of how the Galaxy and its halo formed.
Furthermore, $^6$Li may offer interesting clues to
some fundamental but currently unresolved issues in cosmology and structure formation
on sub-galactic scales.
\end{small}
\end{quotation}
]


\bigskip

\section{Introduction}
Apart from $^7$Li, the bulk of the light elements Li, Be and B
are believed to arise from nonthermal 
nuclear reactions induced by cosmic rays (CRs) 
(Meneguzzi , Audouze  \& Reeves 1971).
In the last decade, extensive observations of LiBeB in population II,
metal-poor halo stars (MPHS) have turned up new and unexpected results, spurring 
controversy as to what type of CR sources and production mechanisms were 
operating in the halo of the early, forming Galaxy (e.g. 
Vangioni-Flam, Cass\'e \& Audouze 2000).
To date, most models of light element evolution in the early Galaxy have 
focused on strong shocks driven by supernovae (SNe) as the principal
sources of CRs.
A general consensus is that 
Be and B in MPHS mainly originate from the ``inverse'' spallation process, 
whereby CR CNO particles are transformed in flight into LiBeB by impinging 
on ISM H or He atoms (Duncan, Lambert \& Lemke 1992).
This can be realized if a sizable fraction of the CRs responsible for 
spallation comprise fresh, CNO-rich SN ejecta
(e.g. 
Vangioni-Flam et al. 2000, 
Ramaty et al. 2000, 
Suzuki, Yoshii \& Kajino 1999, 
Suzuki \& Yoshii 2001,hereafter SY), as opposed to CRs injected from the 
average ISM (e.g. Fields \& Olive 1999).

The origin of $^6$Li in MPHS
(Hobbs 2000 and references therein) is more mysterious, as 
current models involving SN CRs face some difficulties.
A peculiar aspect of Li is that in addition to spallation, the fusion 
process of CR $\alpha$ particles with ambient He atoms can be effective, 
and should actually dominate Li production at low metallicities.
(Note that while both $^7$Li and $^6$Li are synthesized in comparable 
amounts, the CR-produced $^7$Li component is generally overwhelmed by 
the ``Spite plateau'' from primordial nucleosynthesis in the metallicity 
range under consideration; e.g. Ryan et al. 2001.)
If the CR energy spectrum is taken to be a standard power-law distribution 
in momentum (\S 3), one requires a CR injection efficiency much higher than 
normally inferred to reproduce the $^6$Li observations, whether the CR 
composition is metal-enriched or not (Ramaty et al. 2000; SY, see their fig.2).
This raises the question of whether there may have been other sources for
$^6$Li. 

Suzuki \& Inoue (2002; hereafter SI) proposed a new and more natural $^6$Li 
production scenario based on a previously unconsidered CR source: 
CRs accelerated at virialization shocks (VSs), i.e. shocks driven 
by the gravitational infall and merging of sub-Galactic gas clumps during the hierarchical 
build-up of structure in the early Galaxy.
Such shocks are inevitable consequences in the currently standard theory 
of hierarchical structure formation (see however \S 5).
We further discuss a number of testable predictions and important implications 
expected in our scenario for understanding the formation of our Galaxy.
The potential relevance to some fundamental issues
in cosmology and structure formation on sub-galactic scales is also mentioned.

\section{Cosmic Ray Sources in the Early Galaxy}
SNe are known to release $E_{\rm SN} \sim 10^{51}$ erg of kinetic energy in
each explosion, driving strong shocks into the ambient medium.
These SN shocks are favorable sites for efficient CR acceleration through 
the first order Fermi mechanism. 
A plausible value for the injection efficiency $\xi_{\rm SN}$, i.e. 
the fraction of the SN kinetic energy imparted to CRs, is $10-20\%$, deduced 
from comparison of the SN rate and the energy content of CRs currently 
observed in the Galactic disk. 
The global energetics of SNe for the early Galactic halo can be estimated from 
the cosmic star formation history or the total amount of heavy elements 
ejected by halo SNe (see SI for details of the estimation). Both methods 
give $\epsilon_{\rm SN} \sim {\cal E}_{\rm SN} \mu m_p/M_g \simeq 0.15$ keV per
particle as the average specific energy input from SNe.

SNe are not the only sources of mechanical energy (and hence CRs through 
shock acceleration) that may have been active in the early Galaxy.
In the framework of the currently successful picture of hierarchical 
structure formation in the universe, large scale objects are formed through 
the merging and virialization of smaller subsystems, driven by gravitational 
forces acting on the dark matter.
For each merging hierarchy, shocks should inevitably arise in the associated 
baryonic gas component, whereby the kinetic energy of infall is dissipated 
and the gas heated to the virial temperature of the merged halo (e.g. 
White \& Rees 1978). 
It is quite plausible that such virialization shocks (VSs) also 
accelerate CRs (e.g. Miniati et al. 2001).

A guide to how structure formation may have proceeded in our Galaxy, 
particularly for the Galactic halo, may be offered by the recent numerical 
simulations of Galaxy formation by Bekki \& Chiba (2001, hereafter BC;
see also Samland \& Gerhard 2003, Abadi et al. 2003).
It is expected that sub-Galactic structure eventually merge into a single entity in the 
central region at redshift $z \sim 2$, 
whereby the majority of the infall kinetic energy is virialized.
The total energy dissipated at this main VS can be 
evaluated from the virial temperature of the merged system.
The virial temperature $T_v$ for a halo of total mass $M_t$ virializing at
redshift $z \sim 2$ is
$k_B T_v  = \mu m_p G M_t /2 r_v \simeq 0.26 {\rm keV}
\left(M_t / 3 \times 10^{12} {\rm M_\odot}\right)^{2/3}$,
where $r_v$ is the virial radius,
and we have assumed a cosmology with $\Omega_m=0.3$, $\Omega_{\Lambda}=0.7$ and $h=0.7$
(see SI).
The specific energy dissipated at the VS is thus 
$\epsilon_{\rm VS} \sim 3k_B T_v/2 \simeq 0.4$ keV per particle, higher 
than the above estimate for SNe by a factor of $\simeq 2.6$, if we adopt
$M_t = 3 \times 10^{12} {\rm M_\odot}$ (Sakamoto, Chiba \& Beers 2003).
The CR contribution from VSs compared to SNe should similarly be higher, 
as the CR injection efficiency $\xi_{\rm VS}$ should not be 
too different from that for SNe (Miniati et al. 2001).
CRs accelerated by VSs should therefore be at least as 
important as SN CRs, and may well dominate at early epochs.

Another, more speculative but potentially interesting possibility
is outflows powered by massive black 
hole(s), which may have been active in the early Galaxy.
However, there are numerous ambiguities with such a picture,
and here we will concentrate on CRs from VSs and SNe (see SI for more discussion).

\section{Model}
We employ assumptions and parameters deemed most plausible for the SN CRs.
The CR injection efficiency is taken to be $\xi_{\rm SN}=0.15$
for each SN of kinetic energy $E_{\rm SN}=10^{51} {\rm erg}$.
A standard, single power-law distribution in particle momentum
is adopted for the CR spectrum (see SI). 
The injection spectral index is chosen to be $\gamma_{\rm SN}=2.1$,
appropriate for strong SN shocks,
and consistent with the source spectrum inferred for present-day CRs.
As assumed in SY,
the composition of SN CRs is a mixture
of SN ejecta containing freshly synthesized CNO and Fe
and the ambient ISM swept up by the SN blastwave.

CRs from VSs are markedly different from SN CRs in a number of
important ways.
First and foremost, VSs do not synthesize fresh CNO nor Fe,
so that the composition of these CRs
is completely ascertained by the pre-existing ISM.
When the ISM is metal-poor,
these shocks induce very little Be or B production through inverse spallation,
and are only efficient at spawning Li via $\alpha-\alpha$ fusion.
Second, VSs are not necessarily strong ones,
particularly for major mergers of systems with comparable masses
(Miniati et al. 2001),
provided that the pre-shock gas has not cooled significantly.
Shock acceleration should then lead to 
injection indices $\gamma_{\rm VS}$ steeper than the strong shock limit
value of 2,
which works in favor of Li production (\S 4).
Major merger shocks may possess Mach numbers as low as $\simeq 2-3$,
corresponding to $\gamma_{\rm VS} \simeq 2.5-3.3$;
$\gamma_{\rm VS}=3$ is generally chosen below.
As with SN CRs,
we take the spectral shape to be a momentum power-law distribution
and the injection efficiency to be $\xi_{\rm VS}=0.15$.

A further distinction from SNe is that
the VS CR flux should not entail any direct dependence on the metallicity,
which is in fact an obstacle to predictive modeling.
While the hierarchical growth of structure
with respect to cosmic time 
may be evaluated in concrete ways using e.g. extended Press-Schechter formalisms,
relating this to [Fe/H] requires additional knowledge
of how the combination of star formation, SN nucleosynthesis and chemical evolution
in our Galaxy proceeded with cosmic time. 
This involves large uncertainties,
and is not specified in our chemical evolution model.
As a first step,
we choose to describe the time evolution of VS CRs in a simple,
parameterized way,
assuming a `step function' behavior:
VS CRs begin to be injected from a certain time $t_{\rm VS}$,
maintains a constant flux for a duration $\tau_{\rm VS}$,
and then returns to zero.
We take the injection duration $\tau_{\rm VS}$ to be
roughly the dynamical time of the major merger, $\simeq 3 \times 10^8$ yr.
The injection is also assumed to be uniform,
since the effect of the main VS should be global
throughout the gas under consideration, unlike SNe.
The injected VS CR flux integrated over $\tau_{\rm VS}$
is normalized to the above values for $\xi_{\rm VS}$ and $\epsilon_{\rm VS}$.
The true evolutionary behavior of VS activity relative to metallicity 
should actually be probed through future observations of $^6$Li at low [Fe/H] (\S 4, \S 5).

After injection by either SNe or VSs,
the spectral flux $F_i(E,t)$ for each CR element $i$
evolves with time during subsequent interstellar propagation.
This is obtained from time-dependent solutions
of the CR transport equation for a leaky box propagation model (SY).
Using the transported spectra,
we calculate the CR production of LiBeB in the ISM 
including all three types of reactions:
forward spallation of ISM CNO by CR protons and $\alpha$'s,
inverse spallation of CR CNOs by ISM H and He,
and the fusion of CR $\alpha$'s with ISM He.
For more details, consult SY.
We do not consider here
the potentially complicating effects of stellar depletion,
which are highly uncertain at the moment. 

\section{Results}
\begin{figure}[htb]
     \psfig{file=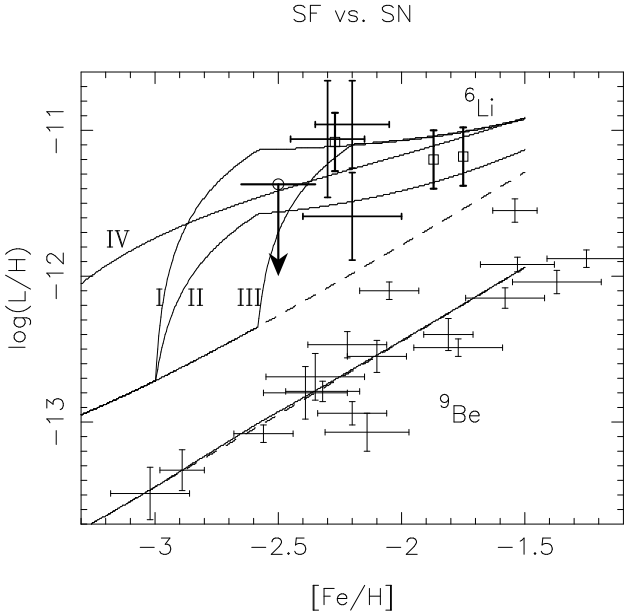,width=8.25cm} 
   \caption{Model results of $^6$Li/H (thick) and Be/H (thin) vs. [Fe/H],
for SN CRs only (dashed curves) and SN plus VS CRs (solid curves),
each label corresponding to the parameter set described in the text.
Also plotted are current observational data
for $^6$Li (thick crosses from Smith et al. 1998 and Nissen et al.2000; 
squares with errors from Asplund et al. 2001; circle with upper limit from 
Aoki et al. 2003) and Be (thin crosses from Boesgaard et al. 1999).
     }
 \label{fig:libe}            
\end{figure}

For our calculations, we have selected the following sets of parameters
for $t_{\rm VS}$, $\tau_{\rm VS}$ and $\gamma_{\rm VS}$, respectively,
labeled models I - IV:
I (0.22, 0.1, 3), II (0.22, 0.1, 2), III (0.32, 0.1, 3), and
IV (0.1, 0.5, 3), where $t_{\rm VS}$ and $\tau_{\rm VS}$ are in units of Gyr.
These were chosen to provide results
exemplary of light element production by VS CRs, in contradistinction to
that by SN CRs.
The evolution of $^6$Li and Be vs. metallicity
calculated for each model until the end of halo chemical evolution
([Fe/H]$\simeq -1.5$)
is shown in Fig.\ref{fig:libe},
along with the current observational data in MPHS.

We discuss some salient points regarding these results.
First, it is confirmed that with our fiducial parameters,
production by SN CRs alone (dashed) 
works very well for the observed Be (and B, not shown),
yet falls short of the observed $^6$Li.
In contrast, with reasonable values for
$\epsilon_{\rm VS}$, $\xi_{\rm VS}$ and $\gamma_{\rm VS}$,
production by VS CRs is capable of explaining the current $^6$Li data
quite adequately.
This mainly owes to two facts:
1) VSs are more energetic than (or at least as energetic as)
SN shocks, as estimated in \S 2, and
2) VS CRs can generate $^6$Li at early epochs independently of the metallicity.
Regardless of the early evolutionary behavior,
identical $^6$Li abundances are attained at the end of the halo phase
for a given $\gamma_{\rm VS}$ (I, III, \& IV),
since this is determined by the time-integrated CR flux,
for which we had assumed a fixed value.
Compared to a flat spectral index of $\gamma_{\rm VS}=2$ (II),
a steeper one of $\gamma_{\rm VS}=3$, more appropriate for low Mach number
VS shocks (\S 3),
results in a larger $^6$Li yield, by about factor of 3.
This is because with a constant total CR energy,
a steeper index implies a larger CR flux
in the subrelativistic energy range $E \sim 10$ -- $100$ MeV,
where the $^6$Li production cross section peaks (see fig.1 in SY).
Note that a conservatively lower specific energy for VSs,
$\epsilon_{\rm VS} \sim 0.15$ keV/particle (i.e. comparable to SNe),
can still be consistent with the available $^6$Li data
provided that $\gamma_{\rm VS} \simeq 3$.

As already stressed, 2) is a consequence of
$^6$Li synthesis being dominated by $\alpha-\alpha$ fusion,
and VSs not creating any new Fe.
Depending on the onset time and duration of the VS,
the $^6$Li abundance may potentially reach large values
quickly at very low metallicity,
which can be followed by a plateau or a very slow rise.
On the other hand, SNe unavoidably give forth to freshly synthesized Fe,
so a correlation with Fe/H must arise;
in fact $^6$Li/H vs. Fe/H for SN CRs can never be much flatter than linear
(e.g. 
Ramaty et al. 2000).
Moreover, since VSs do not eject fresh CNO,
they produce very little Be or B through spallation;
only a minuscule contribution can appear as the ISM becomes metal-enriched.
This may allow an extremely large $^6$Li/Be ratio at low [Fe/H].
Conversely, we see that SN CRs must play an indispensable role
in generating the Be and B observed in MPHS.

Independent of the particular evolutionary parameters,
the following abundance trends are characteristic of VS CR production
and should serve as distinguishing properties of the scenario for future
observations.
Going from high to low metallicity:
a plateau or a very slow decrease in $\log ^6$Li/H vs. [Fe/H],
followed by a steeper decline in some range of [Fe/H]
corresponding to the main epoch of VS;
a steady increase in $^6$Li/Be, possibly up to values exceeding $\simeq 100$,
also followed by a downturn.
These traits are very distinctive and not expected in SN CR models,
for which the slope of $\log ^6$Li/H - [Fe/H] must be $\simeq 1$ or greater,
and the $^6$Li/Be ratio constant at sufficiently low [Fe/H].
Distinction from any production processes in the early universe (e.g.
Jedamzik 2000)
should also be straightforward,
as they predict a true plateau down to the lowest [Fe/H],
in contrast to an eventual decrease for VS CR models.
Further, unique diagnostic features are discussed in \S 5.

\section{$^6$Li as Fossil Record of Dissipative Processes during Galaxy
Formation}

A truly unique and intriguing aspect of the VS CR picture is that
$^6$Li in MPHS can be interpreted and utilized as a fossil record
of dissipative gas dynamical processes in the early Galaxy.
Of particular interest are various correlations expected
between the $^6$Li abundance and the kinematic properties of the stars.
On the one hand,
$^6$Li arises as a consequence of gaseous dissipation through
gravitationally-driven shocks,
and survives to this day as signatures
of the dynamical history of hierarchical structure formation in the early Galaxy.
On the other,
the kinematic characteristics of stars presently observed should reflect
the past dynamical state of their parent gas systems,
because once stars form, they become collisionless
and have long timescales for phase space mixing (e.g. Chiba \& Beers 2000).
Interesting relationships may then exist among the two observables.

For example,
recent intensive studies of the structure and kinematics of MPHS in our Galaxy
based on Hipparcos data (Chiba \& Beers 2000) 
have elucidated the detailed characteristics of our Galaxy's halo,
such as its two-component nature:
an inner halo which is flattened and rotating,
and an outer halo which is spherical and non-rotating.
This dichotomy has been suggested to result from
differences in the physical processes responsible for their formation,
dissipative gas dynamics being crucial for the former,
and dissipationless stellar dynamics determining the latter.
The numerical simulations of Galaxy formation by BC support this conjecture:
the outer halo forms through dissipationless merging of small sub-Galactic clumps
that have already turned into stars (c.f. Searle \& Zinn, 1978),
whereas the inner halo mainly forms through dissipative merging and
accretion of larger clumps
that are still gas rich (c.f. Eggen, Lynden-Bell \& Sandage, 1962).
In our scenario,
$^6$Li production is a direct outcome
of the principal gas dissipation mechanism of gravitational shock heating.
If the above inferences on the formation of halo structure are correct,
$^6$Li should be systematically more abundant
in stars belonging to the inner halo compared to those of the outer halo,
a clearly testable prediction.
An important prospect is that
$^6$Li may provide a quantitative measure
of the effectiveness of gas dynamical processes during formation of halo structure,
rather than just the qualitative deductions allowed by kinematic studies.

Another possibility regards the main epoch of VS with respect to
metallicity (\S 3).
As already mentioned,
the relation between $^6$Li/H and Fe/H
should mirror the time evolution of dissipative energy release through VSs,
but is complicated by being convolved with
the uncertain ingredients of star formation and chemical evolution.
The Fe abundance can be a bad tracer of time, especially at low [Fe/H]
where the effects of dispersion in SN yields can be extremely large
(SY).
Stellar kinematics information may offer a handle on this problem.
The observed relation between [Fe/H] and $<V_\phi>$, the mean azimuthal
rotation velocity of MPHS, seems to manifest a distinctive kink
around [Fe/H] $\sim$ -2 (Chiba \& Beers 2000).
Through chemo-dynamical modeling of the early Galaxy,
BC have proposed that this kink may correspond to the epoch of the major merger (\S 3).
If this was true, a simple expectation in the context of the VS CR model
is that $^6$Li/H should be just rising near this value of [Fe/H],
which is the range occupied by the currently $^6$Li-detected stars;
also expected are a steep decline at lower [Fe/H],
as well as a plateau or slow rise at higher [Fe/H]
(i.e. close to model III in fig.\ref{fig:libe}).
However, any inferences related to [Fe/H]
are always subject to the chemical evolution ambiguities.
A more reliable and quantitative answer may be achieved
by looking for correlations between $^6$Li/H and $<V_\phi>$ without
recourse to [Fe/H],
as $^6$Li is a direct and pure indicator of dynamical evolution in the
early Galaxy.

Thus the VS CR model for $^6$Li bears important implications
for understanding how our Galaxy formed.
If the above mentioned trends are indeed observed,
it would not only confirm the virialization shock origin of $^6$Li,
but may potentially point to new studies of ``$^6$Li Galactic archaeology'',
whereby extensive observations of $^6$Li in MPHS
coupled with detailed chemo-dynamical models
can be exploited as a robust and clear-cut probe
of dissipative dynamical processes that were essential for the formation
of the Galaxy.

Furthermore, $^6$Li in Galactic MPHSs may potentially offer
interesting clues to a number of outstanding current problems
in cosmology and structure formation theory,
all involving physics on sub-galactic scales.
1) The global importance of dynamical feedback by SNe
has been a long-standing uncertainty in galaxy formation theory
(White \& Rees 1978, Abadi et al. 2003).
Further observations of $^6$Li vs. Fe or possibly Be and B at low metallicity
and comparison with detailed theoretical models
may constrain this crucial unknown.
2) The efficiency and location of VSs as commonly assumed on sub-galactic scales
has recently been brought into question (Katz et al. 2002, Birnboim \& Dekel 2003),
with radical implications for how stars and galaxies form
and how the galaxy luminosity function is shaped (Binney 2003).
The early evolution of $^6$Li may provide a direct probe
of this presently speculative but important suggestion.
3) Comparison of theoretical simulations with observations of dwarf galaxy cores
and of satellite galaxies of our Galaxy and within the Local Group
indicate that standard cold dark matter (CDM) produces
much more substructure than is actually seen.
This ``CDM crisis'' may point to non-standard dark matter properties,
such as warm dark matter, self-interacting dark matter or even more exotic proposals
(e.g. Ostriker \& Steinhardt 2003, Madau \& Kuhlen 2003).
On the other hand, the apparent discrepancy may be
the result of strong feedback by SNe or a UV background.
These different possibilities should result in differences
in the early evolution of $^6$Li, potentially constituting a unique probe.
4) From a combined analysis of measurements of cosmic microwave background anisotropies by {\it WMAP}
and of the power spectrum on galactic and sub-galactic scales, it has been suggested
that the spectrum of primordial density fluctuations deviates from a standard, single power-law
(Spergel et al. 2003).
Although less drastic than non-standard dark matter,
this would also modify the growth of structure on subgalactic scales (Madau \& Kuhlen 2003),
and the consequent $^6$Li production at early epochs.
More detailed investigations of these intriguing prospects are certainly necessary,
but in principle, $^6$Li in our Galactic halo may shed light
on these issues of paramount importance for cosmology and the physics of the early Universe.


\section*{References}
\reference
Abadi, M. G. et al.  2003, \apj, 591, 499
\reference 
Aoki, W. et al. 
2003, \aap, submitted
\reference 
Asplund, M. et al. 
2001, in ``Cosmic Evolution'' Eds. Vangioni-Flam et al., 
World Scientific
\reference 
   Bekki, K. \& Chiba, M.  2001, \apj, 558, 666 (BC)
\reference
   Binney, J.  2003, MNRAS, submitted (astro-ph/0308172)
\reference
   Birnboim, Y., \& Dekel, A.  2003, MNRAS, in press (astro-ph/0302161)
\reference 
\reference 
   Boesgaard et al. 
1999, AJ, 117, 1549
\reference 
   Chiba, M. \& Beers, T. C.  2000, \aj, 119, 2843
\reference 
   Duncan, D. K., Lambert, D. L., Lemke, M.  1992, \apj, 401, 584
\reference 
   Eggen, O. J., Lynden-Bell, D. \& Sandage, A. R.  1962, \apj, 136, 748
\reference 
   Fields, B. D. \& Olive, K. A.  1999, New Ast., 4, 255
\reference 
   Hobbs, L. M.  2000, \physrep, 333, 449
\reference 
   Jedamzik, K.  2000, \prl, 84, 3248
\reference
   Katz, N. et al.  2002, astro-ph/0209279
\reference
   Madau, P. \& Kuhlen, M.  2003, astro-ph/0303584
\reference 
   Meneguzzi, M., Audouze, J. \& Reeves, H.  1971, \aap, 15, 337
\reference 
   Miniati, F. et al. 
2001, \apj, 559, 59
\reference 
   Nissen, P. E. et al. 
2000, \aap, 357, L49
\reference
   Ostriker, J. P. \& Steinhardt, P.  2003, Science, 300, 190
\reference 
   Ramaty, R. et al. 
2000, \apj, 534, 747

\reference 
   Ryan, S. G. et al. 
2001, \apj, 549, 55

\reference 
   Sakamoto, T., Chiba, M. \& Beers, T. C.  2003, \aap, 397, 899
\reference
   Samland, M. \& Gerhard, O. E.  2003, \aap, 399, 961
\reference 
   Searle, L. \& Zinn, R.  1978, \apj, 225, 357
\reference 
   Smith, V. V., Lambert, D. L., \& Nissen, P. E.  1998, \apj, 506, 405
\reference
   Spergel, D. N. et al.  2003, \apjs, 148, 175
\reference 
   Suzuki, T. K., Yoshii, Y. \& Kajino, T.  1999, \apj, 522, L125
\reference 
   Suzuki, T. K. \& Yoshii, Y.  2001, \apj, 549, 303 (SY)
\reference 
Suzuki, T. K. \& Inoue, S. 2002, ApJ, 573, 168
\reference 
   Vangioni-Flam, E., Cass\'e, M. \& Audouze, J.  2000, \physrep, 333, 365
\reference 
   White, S. D. M. \& Rees, M. J.  1978, \mnras, 183, 341


\end{document}